\documentclass{aa} 
\usepackage{txfonts}
\usepackage{longtable}  
\usepackage{rotating}
\usepackage{natbib}
\usepackage{graphicx}
\usepackage{graphics}
\usepackage{psfrag}
\usepackage{amssymb}
\bibpunct{(}{)}{;}{a}{}{,}
\def\Teff{\ensuremath{T_{\mathrm{eff}}}}
\def\logg{\ensuremath{\log g}}

\def\vsini{\ensuremath{{\upsilon}\sin i}}
\def\kms{$\mathrm{km\,s}^{-1}$}

\def\llm{{\sc LLmodels}}

\def\logl{\ensuremath{\log L/L_{\odot}}}

\def\M{\ensuremath{M_{\odot}}}

\def\synth{{\sc SYNTH3}}
\def\logR{\ensuremath{\log R^{'}_{HK}}}
\begin{document} 
\title{Detection of a magnetic field in three old and inactive solar-like planet-hosting stars\thanks{Based on observations made with ESO Telescopes at the La Silla Paranal Observatory under programme ID 089.D-0302.}} 
\subtitle{} 
\author{L. Fossati\inst{1,2}     \and
        O. Kochukhov\inst{3}     \and 
	J.~S. Jenkins\inst{4}    \and
	R.~J. Stancliffe\inst{1} \and
	C.~A. Haswell\inst{2}    \and
	A. Elmasli\inst{5}       \and
	E. Nickson\inst{2}
}
\institute{
	Argelander-Institut f\"ur Astronomie der Universit\"at Bonn, Auf dem 			H\"ugel 71, 53121, Bonn, Germany\\
	\email{lfossati@astro.uni-bonn.de} 
	\and
	Department of Physical Sciences, Open University, Walton Hall, Milton 			Keynes, MK7 6AA, UK
	\and
	Department of Physics and Astronomy, Uppsala University, 751 20, 			Uppsala, Sweden 
	\and
	Departamento de Astronomia, Facultad de Ciencias Fisicas y 				Matematicas, Universidad de Chile, Camino El Observatorio \# 1515, Las 			Condes, Santiago, Chile
	\and
	Astronomy and Space Sciences Department, Ankara University, Tandogan, 			06100, Ankara, Turkey
} 
\date{} 
\abstract
{}
{Our understanding of magnetic fields in late-type stars is strongly driven by what we know of the solar magnetic field. For this reason, it is crucial to understand how typical the solar dynamo is. To do this we need to compare the solar magnetic field with that of other stars as similar to the Sun as possible, both in stellar parameters and age, hence activity. We present here the detection of a magnetic field in three planet-hosting solar-like stars having a mass, age, and activity level comparable to that of the Sun.}
{We used the HARPSpol spectropolarimeter to obtain high-resolution high-quality circularly polarised spectra of HD\,70642, HD\,117207, and HD\,154088, using the Least-Squares Deconvolution technique to detect the magnetic field. From the Stokes $I$ spectra, we calculated the \logR\ activity index for each star. We compared the position of the stars in the Hertzsprung-Russell diagram to evolutionary tracks, to estimate their mass and age. We used the lithium abundance, derived from the Stokes $I$ spectra, to further constrain the ages.}
{We obtained a definite magnetic field detection for both HD\,70642 and HD\,154088, while for HD\,117207 we obtained a marginal detection. Due to the lower signal-to-noise ratio of the observations, we were unable to detect the magnetic field in the second set of observations available for HD\,117207 and HD\,154088. On the basis of effective temperature, mass, age, and activity level the three stars can be considered solar analogs.}
{HD\,70642, HD\,117207, and HD\,154088 are ideal targets for a comparative study between the solar magnetic field and that of solar analogs.}
\titlerunning{Detection of a magnetic field in three old and inactive solar-like}
\authorrunning{L. Fossati et al.}
\maketitle
\section{Introduction}\label{introduction}
Our knowledge about magnetic fields in late-type stars is strongly driven by what we know of the Sun; but how typical is the solar magnetic field? For example, the Sun's magnetic field and activity cycle might be rather peculiar due to the presence of planets, Jupiter in particular \citep{zaq1997,wilson2008}, which has an orbital period consistent with the periodicity of the solar cycle.  One way to look for an answer to ``how typical is the solar dynamo?'' is to look for and characterise the magnetic field of stars as massive, old and inactive as the Sun.

\citet{petit2008} determined the geometry of the magnetic field for four solar-like stars, concluding that slowly rotating (inactive) stars tend to host a stronger poloidal component of the magnetic field compared to the toroidal one, with the latter increasing its strength at the expense of the former with increasing rotation (activity). Of the four stars analysed by \citet{petit2008}, only \object{18\,Sco} (HD\,146233) is as old and inactive as the Sun, while the other three stars are progressively younger and therefore more active. The magnetic field geometry of 18\,Sco is similar to that of the Sun, but one comparison star is not statistically significant. In general, any such comparison should also aim to sample over the stellar activity cycle.

In April 2012 we observed 29 planet-hosting stars aiming to look for and study their magnetic fields and possible connection to star-planet interaction. Most of the stars for which we detected a magnetic field were either young and active or evolved away from the main sequence (Hertzsprung-gap stars). The results of this program will be published in a separate paper. Here we present the detection of a magnetic field in three old and inactive solar-like planet-hosting stars: \object{HD\,70642}, \object{HD\,117207}, \object{HD\,154088}. Together with 18\,Sco, these objects will ultimately allow us to understand how typical is the solar dynamo, as well as detecting a connection between the stellar activity cycle and giant planets, if any.
\section{Observations and magnetic field detection}\label{observations}
We observed the three stars with the HARPSpol polarimeter
\citep{snik2011,piskunov2011} feeding the HARPS spectrograph \citep{mayor2003} attached to the ESO 3.6-m telescope in La Silla, Chile. The observations, covering the 3780--6910\,\AA\ wavelength range with a spectral resolution $R\sim$115\,000, were obtained using the circular polarisation analyser. We observed each star with a sequence of four sub-exposures obtained rotating the quarter-wave retarder plate by 90$^\circ$ after each exposure, i.e. 45$^\circ$, 135$^\circ$, 225$^\circ$ and 315$^\circ$. The exposure times have been set according to the stellar brightness and sky conditions; the resulting signal-to-noise ratio (S/N) is listed in Table~\ref{tab:Bfield}. The observing log is given in Table~\ref{obs_log}. 
\begin{table}[h!]
\caption[]{Basic datas of the observations.}
\label{obs_log}
\begin{center}                     
\begin{tabular}{lccccccr}
\hline
\hline
Star  	& $V$	& Date        & MJD    & Exp Time      \\
Name 	& 	& 04/2012     &	       & [s]	       \\
\hline
HD\,70642 		& 7.17 	& 17 &  56034.9675      & 4$\times$1160 \\
HD\,117207		& 7.26 	& 17 &  56035.1746      & 4$\times$1400 \\
			&  	& 21 &  56039.0703      & 4$\times$1450 \\ 	     
HD\,154088		& 6.58	& 17 &  56035.3896      & 4$\times$670  \\
			&  	& 21 &  56039.4035      & 4$\times$470  \\ 	     
\hline
\end{tabular}
\end{center}                     
\tablefoot{The date, given in column three, corresponds to the night of observation in April 2012. The modified Julian date (MJD) is that of the beginning of the observation.}
\end{table}

We reduced and calibrated the data with the {\sc REDUCE} package \citep{piskunov2002}, obtaining one-dimensional spectra which were combined using the ``ratio'' method in the way described by \citep{bagnulo2009}. We then re-normalised all spectra to the intensity of the continuum obtaining a spectrum of Stokes $I$ ($I/I_c$) and $V$ ($V/I_c$), plus a spectrum of the diagnostic null profile \citep[$N$ - see][]{bagnulo2009}, with the corresponding uncertainties.

The magnetic field detection technique used in this work requires the knowledge of the atmospheric parameters (effective temperature - \Teff\ and surface gravity - \logg) and metallicity ([Fe/H]). We extracted from the literature a number of independent photometric \citep[Infrared Flux Method - ][]{casagrande2011,ramirez2005} and spectroscopic \citep{prieto1999,santos2004,valenti2005,taylor2005,bond2006,sousa2008,ghezzi2010a,gonzalez2010} determinations and set \Teff, \logg, [Fe/H], and their uncertainties, as the average values and their standard deviation. The final adopted set of parameters and metallicity is given in Table~\ref{parameters1}. Averaging results obtained with different techniques (i.e. photometric and spectroscopic) allows one to reduce systematic uncertainties \citep{fossati2011}. 
\begin{table}[h!]
\caption[]{Adopted atmospheric parameters compared to that of the Sun.}
\label{parameters1}
\begin{small}
\begin{center}                     
\begin{tabular}{lcccc}
\hline
\hline
Star & \Teff & \logg  & [Fe/H]  & \vsini  \\
Name &   [K] & [cgs]  &         & [\kms]  \\
\hline
HD\,70642   & 5683(42/10) & 4.41(03/9) & $+$0.17(04/9) &  0.3 \\
HD\,117207  & 5678(56/6)  & 4.32(13/6) & $+$0.22(05/5) &  1.0 \\
HD\,154088  & 5423(51/4)  & 4.40(11/4) & $+$0.31(03/3) &  1.9 \\
\hline
Sun         & 5777 & 4.44 & 0.00 &  1.2 \\
\hline
\end{tabular}
\end{center}                     
\end{small}
\tablefoot{The numbers given in parenthesis for \Teff, \logg, and [Fe/H] correspond to the uncertainty and adopted number of sources (see Sect.~\ref{li-age-logR}). The projected rotational velocity is taken from \citet{valenti2005}.}
\end{table}

In late-type stars, polarisation signatures are usually undetectable in individual spectral lines in both Stokes $I$ and $V$, unless in the presence of a strong field and with exceptionally good data \citep{donati1997}. To detect magnetic fields we used the Least-Squares Deconvolution technique \citep[LSD;][]{donati1997}, which combines line profiles (assumed to be identical) centered at the position of the individual lines and scaled according to the line strength and sensitivity to a magnetic field. The resulting mean profiles ($I$, $V$, and $N$) were obtained by combining a few thousand spectral lines with a great increase in S/N and therefore sensitivity to polarisation signatures. We computed the LSD profiles of Stokes $I$, $V$, and of the null profile using the methodology and the code described in \citet{kochukhov2010}. We prepared the line mask used by the LSD code separately for each star adopting the stellar parameters listed in Table~\ref{parameters1}. We extracted the line parameters from the Vienna Atomic Line Database \citep[VALD;][]{vald1,vald2,vald3}. For each star we used about 3000 metal lines, all stronger than 30\% of the continuum, avoiding lines with extended wings (e.g. \ion{Mg}{i}\,b and hydrogen lines) and in spectral regions affected by the presence of telluric lines.

\begin{figure*}[ht!]
\sidecaption
\includegraphics[width=127mm,clip]{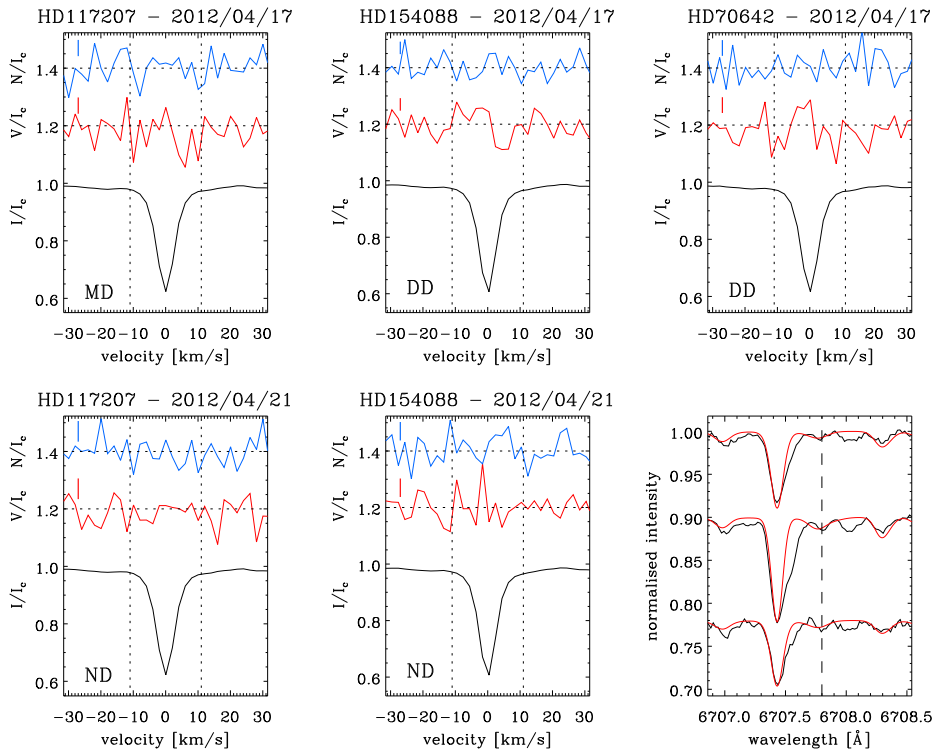}
\caption{$I$, $V$ and $N$ LSD profiles (from bottom to top) obtained for HD\,117207 (left panels), HD\,154088 (middle panels), and HD\,70642 (top-right panel). The $V$ and $N$ profiles were expanded by a factor of 700 and shifted upwards. The uncertainties are shown as bars at the top-left hand side of each panel. All profiles are shifted to the rest frame. The detection flag is given at the bottom-left corner of each panel. The vertical dashed lines indicate the range adopted for the calculation of the magnetic field. The bottom-right panel shows a comparison between observed and synthetic spectra for HD\,117207 (top), HD\,154088 (middle), and HD\,70642 (bottom) in the region of the \ion{Li}{i} feature at $\sim$6707\,\AA, for which the approximate position is indicated by a vertical dashed line.}
\label{fig:LSDprofiles}
\end{figure*}

Figure~\ref{fig:LSDprofiles} shows the obtained LSD profiles, while Table~\ref{tab:Bfield} gives the results gathered from their analysis. We defined the magnetic field detection making use of the false alarm probability \citep[FAP;][]{donati1992}, considering a profile with FAP\,$<\,10^{-5}$ as a definite detection (DD), $10^{-5}\,<$\,FAP\,$\,<10^{-3}$ as a marginal detection (MD), and FAP\,$>\,10^{-3}$ as a non-detection (ND). To further check the magnetic field detections are not spurious, we calculated the FAP for the null profile in the same velocity range as used for the magnetic field, obtaining ND in all cases. We also calculated the FAP for equivalent velocity ranges displaced both redwards and bluewards to sample the continuum in the Stokes $I$ spectrum. The results for both Stokes $V$ and the null profile are given in column 6 of Table~\ref{tab:Bfield}. The much higher FAP obtained in these tests are consistent with our detections being genuine. In addition, we checked both Stokes $V$ and the null profile are consistent with the expected noise properties (e.g., Stokes $V$ uncertainties consistent with the standard deviation of the null profile).
%
\begin{table*}[]
\caption[]{Results from the LSD analysis.}
\label{tab:Bfield}
\begin{center}                     
\begin{tabular}{lcccccccc}
\hline
\hline
Star & Date & $<$B$_z>(V)$ & FAP~($V$) & Detection & FAP~($V$)            & S/N & S/N	   & \# lines \\
Name &      & [G]          & line      & $V$       & blue cont./red cont. & $I$ & $V_{LSD}$ &	      \\
\hline
HD\,70642   & 17 &    0.54$\pm$0.59 & 9.7$\times$10$^{-6}$ & {\bf DD} & 1.4$\times$10$^{-1}$/5.4$\times$10$^{-1}$ & 339 & 17864 & 2873 \\
HD\,117207  & 17 &    1.36$\pm$0.62 & 1.1$\times$10$^{-4}$ & {\bf MD} & 1.4$\times$10$^{-1}$/1.6$\times$10$^{-1}$ & 389 & 17471 & 2859 \\
            & 21 & $-$0.55$\pm$0.77 & 9.9$\times$10$^{-1}$ &      ND  & 4.8$\times$10$^{-1}$/2.3$\times$10$^{-1}$ & 251 & 13808 & 2859 \\
HD\,154088  & 17 &    2.72$\pm$0.49 & 3.0$\times$10$^{-6}$ & {\bf DD} & 3.3$\times$10$^{-1}$/2.7$\times$10$^{-1}$ & 288 & 21280 & 3299 \\
            & 21 &    0.96$\pm$0.71 & 1.2$\times$10$^{-3}$ &      ND  & 5.1$\times$10$^{-1}$/9.9$\times$10$^{-1}$ & 206 & 14353 & 3299 \\
\hline
Star & Date & $<$B$_z>(N)$ & FAP~($N$) & Detection & FAP~($N$)            & S/N & S/N       & \# lines \\
Name &      & [G]	   & line      & $N$       & blue cont./red cont. & $I$ & $V_{LSD}$ &	       \\
\hline
HD\,70642   & 17 &    0.21$\pm$0.59 & 4.2$\times$10$^{-1}$ &      ND  & 1.8$\times$10$^{-1}$/1.2$\times$10$^{-2}$ & 339 & 17864 & 2873 \\
HD\,117207  & 17 & $-$0.05$\pm$0.62 & 4.2$\times$10$^{-1}$ &      ND  & 2.9$\times$10$^{-2}$/2.5$\times$10$^{-1}$ & 389 & 17471 & 2859 \\
            & 21 &    0.55$\pm$0.77 & 6.0$\times$10$^{-1}$ &      ND  & 8.1$\times$10$^{-1}$/8.1$\times$10$^{-1}$ & 251 & 13808 & 2859 \\
HD\,154088  & 17 & $-$0.47$\pm$0.79 & 4.0$\times$10$^{-2}$ &      ND  & 1.3$\times$10$^{-1}$/4.6$\times$10$^{-1}$ & 288 & 21280 & 3299 \\
            & 21 & $-$0.34$\pm$0.71 & 5.5$\times$10$^{-1}$ &      ND  & 1.4$\times$10$^{-1}$/5.8$\times$10$^{-1}$ & 206 & 14353 & 3299 \\
\hline
\end{tabular}
\end{center}                     
\tablefoot{The S/N (per-pixel) of Stokes $I$ is that of the observed spectum and it has been calculated over an 0.5\,\AA\ region at $\sim$5060\,\AA. The S/N of Stokes $V$ is that of the LSD profile. Column six lists the FAP calculated in the continuum region bluewards and redwards of the spectral line, over a range as broad as that adopted to derive $<$B$_z>$. The last column lists the number of lines used in the line mask. For all stars we adopted a range of 22\,\kms\ (i.e., $\pm$11\,\kms\ from the line center) for the calculation of the magnetic field.}
\end{table*}

The non-detections resulting from the analysis of the spectra of HD\,117207 and HD\,154088 on the night of April 21st are most likely due to the lower S/N compared to that obtained on the night of April 17th. This was caused by the highly variable seeing on the night of April 21st, which did not allow us to have enough control on the exposure times in order to achieve the desired S/N. 
\section{Lithium abundance, age and activity}\label{li-age-logR}
To find whether the three stars presented here are good comparisons to the Sun, we need to establish their age and activity level as accurately as possible. We estimated the mass and age of the stars by fitting evolutionary tracks to their position in the Hertzsprung-Russell (HR) diagram. We derived luminosities on the basis of the $V$ magnitudes and Hipparcos distances \citep{vanleeuwen2007}, adopting the bolometric correction by \citet{flower1996}. 

\begin{figure}[ht!]
\begin{center}
\includegraphics[width=65mm,clip,angle=-90]{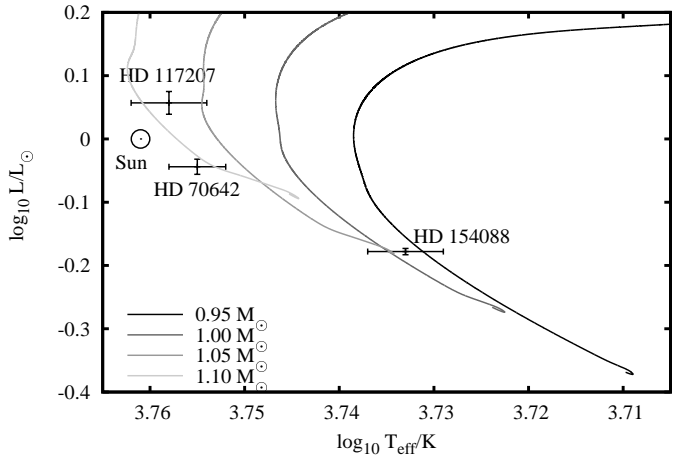}
\caption{Position of HD\,70642, HD\,117207 and HD\,154088 in the HR diagram in comparison with evolutionary tracks calculated with [Fe/H]=0.3\,dex. The position of the Sun is also indicated and its displacement from the 1.00\,\M\ track is due to the non-solar metallicity used for the tracks.}
\label{fig:HR}
\end{center}
\end{figure}

We used the {\sc stars} stellar evolution code \citep{eggleton1971,stancliffe2009} to construct models for stars of between 0.9 to 1.5\,\M\  (with a mass spacing of 0.05\,\M), for metallicities of $Z$=0.01, 0.02 and 0.04. These models have been evolved from the pre-main sequence to the giant branch using 999 meshpoints, a mixing length parameter of 2.0 and without the inclusion of convective overshooting. Figure~\ref{fig:HR} shows the position of the three target stars in the HR diagram, in comparison with {\sc stars} evolutionary tracks calculated with a metallicity of [Fe/H]=0.3\,dex ($Z$=0.04). By taking into account the uncertainties on \Teff, \logl, and [Fe/H], for HD\,70642, HD\,117207 and HD\,154088 we estimated an age of 3$\pm$3, 5$\pm$2, and 5$\pm$3\,Gyr, respectively. The derived stellar masses are listed in Table~\ref{parameters2}.

To further constrain the age, we measured the Li abundance from the feature at $\sim$6707\,\AA, adopting hyperfine structure from \citet{smith1998} and the meteoritic/terrestrial isotopic ratio Li$^6$/Li$^7$\,=\,0.08 \citep{rosman1998}. To measure the Li abundance, we fitted synthetic spectra, calculated adopting the \llm\ \citep{shulyak2004} model atmosphere code and the \synth\ \citep{kochukhov2007} spectral synthesis code, to the observations. The Li abundance obtained for each star is listed in Table~\ref{parameters2}, while the bottom-right panel of Fig.~\ref{fig:LSDprofiles} shows a comparison between observed and synthetic spectra in the region of the analysed \ion{Li}{i} feature, demonstrating the very low Li abundance of the three stars. For both HD\,70642 and HD\,117207 the derived Li abundance is in agreement with that given by \citet{ghezzi2010b} and \citet{gonzalez2010}. A comparison of the Li abundance with results by \citet{sestito2005} allows us to conclude that all three stars are older than at least 3\,Gyr, hence they are as old as or older than the Sun. 

The chromospheric activity $S$-index was measured by comparing the flux in the core of the \ion{Ca}{ii}\,H\&K lines against the flux in the surrounding continuum region, as explained by \citet{jenkins2006,jenkins2008}. We convert this value into the classic \logR\ index using the recipes described in \citet{noyes1984} after conversion to the Mt.\,Wilson system of measurements \citep[see][]{duncan1991}. A more thorough description of the calibration method will be presented in a forthcoming paper (Fossati et al. 2013, in prep). The derived \logR\ for all three stars lie between $-$4.93 and $-$5.05, typical of quiescent solar-like stars on the main sequence \citep{jenkins2011}. In addition, the derived \logR\ activity index is in agreement with that previously obtained by other authors \citep{jenkins2006,wright2004}, showing that our observations have not been taken during a period of excess activity. The higher \logR\ value for HD\,70642 agrees with the slightly younger age found for this star from the isochrone fitting, assuming a rotationally driven magnetic dynamo model that weakens with age. 
\begin{table}[h!]
\caption[]{Derived age range, mass, Li abundance and \logR\ in comparison to that of the quiet Sun.}
\label{parameters2}
\begin{center}                     
\begin{tabular}{lccccccccc}
\hline
\hline
Star & Age   & Mass & $\log n(\rm {Li})$ & \logR \\
Name & [Gyr] & \M   &			 &	 \\
\hline
HD\,70642   & 3--6 & 1.05(10) & $<$0.85 & -4.935(14) \\
HD\,117207  & 3--7 & 1.08(03) & $<$0.85 & -5.043(24) \\
HD\,154088  & 3--8 & 0.97(05) & $<$0.75 & -4.992(40) \\
\hline
Sun         & 4.6  & 1.00     &   1.05  & -4.960 \\
\hline
\end{tabular}
\end{center}                     
\tablefoot{The uncertainties given for mass and \logR\ corresponds to that of the last digits.}
\end{table}

\section{Discussion}\label{discussion}
Except for a slight metal enrichment, HD\,70642 and HD\,117207 have stellar parameters very similar to that of the Sun. The two stars have an age in the range of 3--6\,Gyr for HD\,70642 and 3--7\,Gyr for HD\,117207, proving they are about as old as the Sun, if not older. For both stars, the derived mass and age values are in agreement with previous determinations \citep{valenti2005,takeda2007,ghezzi2010a,gonzalez2010,casagrande2011}. In particular, \citet{casagrande2011} estimated masses and ages of the two stars using two independent stellar evolution codes, obtaining results which are very close to our values. For HD\,117207, \citet{wright2004} reported a rotation period of 36\,days, about 10\,days longer than that of the Sun, further indicating the old age and low activity of the star.

HD\,154088 is slightly cooler and more metal rich, compared to both HD\,70642 and HD\,117207. For this star, we estimated an age range of 3--8\,Gyr, in agreement with previous determinations \citep{valenti2005,takeda2007}. For HD\,154088, \citet{wright2004} measured a rotation period of 42\,days, almost double than that of the Sun.

On the basis of the estimated temperature, mass, age and activity, and following the definition given by \citet{soderblom1998}, HD\,70642, HD\,117207, and HD\,154088 can be considered solar analogs (note that within the uncertainties, the metallicity of HD\,154088 is consistent with the definition of solar analog). Although HD\,117207 would require further confirmation of the presence of a detectable magnetic field, the three stars are ideal targets to characterise their magnetic field geometry and compare it to that of the Sun, to study how typical the solar dynamo is. The cooler temperature of HD\,154088 might not make it such a good comparison for the Sun, but it will allow one to look for differences in the stellar dynamo as a function of mass, for stars of similar ages. In addition, HD\,70642 and HD\,117207 host Jupiter-like planets with Jupiter-like orbital periods ($\sim$6 and 7 years, respectively). It will be very valuable to determine the periodicity of their activity cycle, to check whether it is comparable with the orbital period of the hosted planets.
\begin{acknowledgements}
This work was supported by an STFC RollingGrant (L.F., C.A.H.). E.N. is supported by an STFC studentship. O.K. is a Royal Swedish Academy of Sciences Research Fellow, supported by grants from the Knut and Alice Wallenberg Foundation and the Swedish Research Council. J.J. acknowledges funding by Fondecyt through grant 3110004 and partial support from CATA (PB06, Conicyt), the  GEMINI-CONICYT FUND and from the Comit\'e Mixto ESO-GOBIERNO DE CHILE. RJS is the recipient of a Sofja Kovalevskaja Award from the Alexander von Humboldt Foundation. We thank Stefano Bagnulo for fruitful discussions.
\end{acknowledgements}
\end{document}